\documentclass{svproc}
\usepackage{graphicx}%
\usepackage{multirow}%
\usepackage{amsmath,amssymb,amsfonts}%
\usepackage{mathrsfs}%
\usepackage[title]{appendix}%
\usepackage{xcolor}%
\usepackage{textcomp}%
\usepackage{manyfoot}%
\usepackage{booktabs}%
\usepackage{algorithm}%
\usepackage{algorithmicx}%
\usepackage{algpseudocode}%
\usepackage{listings}%
\usepackage{subeqnarray}
\usepackage{url}

\def\orcidID#1{\unskip$^{[#1]}$} 

\usepackage{xr}
\makeatletter
\newcommand*{\addFileDependency}[1]{
\typeout{(#1)}%
\@addtofilelist{#1}
\IfFileExists{#1}{}{\typeout{No file #1.}}
}\makeatother
\newcommand*{\myexternaldocument}[1]{%
\externaldocument[supp-]{#1}%
\addFileDependency{#1.tex}%
\addFileDependency{#1.aux}%
}
\myexternaldocument{supplementary_materials}

\begin{document}
\mainmatter              
\title{Improved Community Detection using Stochastic Block Models}
\titlerunning{Improving SBM clustering}  
%
\author{Minhyuk Park\inst{1}\orcidID{0000-0002-8676-7565} \and 
Daniel Wang Feng\inst{1}\orcidID{0000-0003-1574-8225} \and
Siya Digra\inst{1}\orcidID{0009-0004-5957-8939} \and The-Anh Vu-Le\inst{1}\orcidID{0000-0002-4480-5535} \and George Chacko\inst{1}\orcidID{0000-0002-2127-1892} \and Tandy Warnow\inst{1}\orcidID{0000-0001-7717-3514
}}
\authorrunning{Park et al.} 

\institute{Siebel School of Computing and Data Science, University of Illinois Urbana-Champaign, Urbana IL 61801\\
\email{\{minhyuk2,warnow,chackoge\}@illinois.edu}\\ 
}

\maketitle              

\begin{abstract} Community detection approaches resolve complex networks into smaller groups (communities) that are expected to be relatively edge-dense and well-connected. The stochastic block model (SBM) is one of several approaches used to uncover community structure in graphs. In this study, we demonstrate that SBM software applied to various real-world and synthetic networks produces poorly-connected to disconnected clusters. We present simple modifications to improve the connectivity of SBM clusters, and show that the modifications improve accuracy using simulated networks.
\end{abstract}
\section{Introduction}\label{sec:introduction}
Community detection is a task in which nodes of a network are partitioned into subsets, each called a community or a cluster (the terms are interchangeable in this manuscript) \cite{Fortunato2022,newman2004detecting}.
Communities do not have to cover an entire network \cite{Miasnikof2023}, and may also be overlapping \cite{jakatdar2022aoc,oslom,xie2013overlapping}.

Many community detection methods are based on optimization criteria that reflect one or more of the following properties:  preference for clusters that are dense  and so have many intra-cluster edges; that are separated from the rest of the network and so have relatively fewer inter-cluster edges; and finally have good cut-conductance, which means they do not have small edge cuts \cite{kannan2004clusterings} and are said to be ``well-connected"
\cite{park2024,traag2019louvain}.

Despite the natural expectation that clusters should be well-connected, this property does not result from some clustering methods \cite{traag2019louvain,zhu2013local}. We have previously reported that the Leiden algorithm, Infomap, Iterative-K-core Clustering (IKC) and Markov Clustering (MCL) community detection algorithms produce clusters that are not well connected \cite{park2024}.
Moreover, \cite{traag2019louvain} documented that the Louvain algorithm can produce disconnected clusters.

Modifying such clusters to improve connectivity is a logical remediation, and 
the Connectivity Modifier \cite{Ramavarapu2024} is one such method that recursively modifies an input clustering to ensure that all final clusters meet  an explicit well-connectedness criterion.

In this study, we examine Stochastic Block Models (SBMs) \cite{karatacs2018application,Lee2019}, widely used for clustering networks. On a collection of more than 100 real-world networks, we find that the  SBM clustering methods in graph-tool \cite{graph-tool} frequently produce disconnected clusters. We explore three techniques for modifying the clustering to improve the connectivity: simply returning the connected components, repeatedly finding and removing small edge cuts until all clusters meet a mild criterion to be considered well-connected, or applying the recursive Connectivity Modifier method to the clustering. 

The rest of this manuscript is as follows. In Section \ref{sec:methods}, we describe networks, clustering methods, and evaluation criteria used in our study. We present the results of our experiments in Section \ref{sec:results}, and our discussion of these results in Section \ref{sec:discussion}. We conclude with a summary of our findings on real world and synthetic networks and outline future directions in Section \ref{sec:conclusion}.

\section{Materials and Methods}\label{sec:methods}

\vspace{-.1in}
\subsection{Networks}

\vspace{-.02in}
\subsubsection{Real-world networks}

 We collected a set of 122 real-world networks that range in size from 11 to 13,989,436 nodes. Of these, 
 120 are from  the Netzschleuder network catalogue  \cite{peixoto2020netzschleuder} and we also include the Orkut network (3,072,441 nodes) and the Curated Exosome Network (13,989,436 nodes) \cite{park2024}.
The Netzschleuder network set includes 10 small networks with at most 1000 nodes, 
  103 medium-sized networks between 1000 and 1,000,000 nodes, and  7 networks with at least 1,000,000  nodes (see Supplementary Materials) \cite{park_sbmcd_suppl_2024} 
  for the full list of networks).
  All real-world networks used in this study were pre-processed to remove self-loops and parallel edges and were treated as unweighted and undirected.

 \vspace{-.1in}
\subsubsection{Synthetic networks} From a previous study \cite{park2024}, we used synthetic networks that were generated using the LFR \cite{lancichinetti2008benchmark} software.
These networks were generated based on parameters obtained from clusterings computed on five real-world networks using the Leiden algorithm \cite{leiden-code} optimizing either the modularity criterion \cite{newman2004finding} or the Constant Potts Model criterion \cite{traag2019louvain} with
 five different resolution values (0.0001, 0.001, 0.01, 0.1, 0.5).
 These LFR  networks range in size from 34,546 nodes to 3,774,768 nodes. As reported in \cite{park2024}, a few  of these LFR networks had a high incidence of disconnected ground-truth clusters and were not suitable for analysis in this study.

\subsection{Stochastic Block Models} \label{sec:methods-sbm}

We used SBM implemented in graph-tool \cite{graph-tool} as a clustering method  with the option of three different models: degree-corrected \cite{dcsbm}, non degree-corrected\cite{holland1983sbm}, and Planted Partition \cite{ppsbm}. We generated networks and ground truth clusterings under all three models using parameters estimated from real-world networks and clusterings given as input to the graph-tool software. For each network, we selected the model that had the best fit--i.e., the one with the lowest description length--as our most preferred  SBM model. We refer to that model as the ``selected SBM", and use it in subsequent post-processing treatments.  Versions and commands for these models are in the Supplementary Materials.

 \vspace{-.1in}
\subsection{Post-processing treatments to improve connectivity}

In this study, if a cluster $c$ has $n_c$ nodes, we consider it to be well-connected when its minimum edge cut size is strictly greater than $\log_{10}(n_c)$, a default used in \cite{park2024}. Else, we consider the cluster to be poorly-connected.

The Connectivity Modifier (CM) \cite{park2024}, implemented as a pipeline \cite{Ramavarapu2024}, is designed to modify clusterings in order to ensure that all clusters are well-connected and that no cluster is too small. 
In prior work \cite{park2024}, we found that the CM pipeline typically improved Leiden clustering accuracy on synthetic networks, and that when it reduced accuracy this was due to removing small clusters.
Hence, in this study, we have eliminated the filtering of small clusters, and restricted the CM pipeline to modifying clusterings in order to ensure edge-connectivity. 

We evaluate the use of this simplified CM approach as well as two other   post-processing treatments, each of which takes as input a clustering $\mathcal{C}$ of a network $N$, and modifies it, if necessary, to ensure some  standard for edge-connectivity. 
The three treatments we study are:
\begin{itemize}
    \item {\bf CC (Connected Components)}: If a cluster is disconnected, i.e., has two or more connected components, we return each of its connected components as a cluster.
    \item {\bf WCC (Well-Connected Clusters}): We modify clusters by repeatedly removing small edge cuts of size at most $\log_{10}(n)$ until each cluster is well-connected.  To find small edge cuts, we  use  VieCut \cite{henzinger2018practical}. Note that removing an edge cut means removing the edges but not the endpoints, and that if we remove a minimum edge cut we break the cluster into exactly two pieces.
    Each of these pieces is then examined for well-connectedness and further processed, if needed.
    \item {\bf CM (Connectivity Modifier)}: Here we apply the inner loop of the Connectivity Modifier pipeline \cite{Ramavarapu2024}. 
    If a cluster $C$ has a small edge cut, then removal of the edge cut divides $C$ into two subsets, and each of these is then ``re-clustered" using the same clustering method used to produce the input clustering $\mathcal{C}$.
    These clusters are then added back into the iterative algorithm, which checks each cluster for being well-connected. Each iteration finds and removes small edges cuts, and then reclusters the two sets. The iteration stops when the cluster satisfies the edge-connectivity criterion.  Note that because we apply only the inner loop of the CM pipeline, we do not filter out small clusters. 
\end{itemize}
Versions and commands for these post-processing treatments are in the Supplementary Materials.

\vspace{-.1in}
\subsection{Evaluation} 
We report cluster  statistics, including  percent of clusters that are connected, percent well-connected, and percent poorly connected.
We also report cluster size distributions and node coverage after restriction to clusters of size at least two.
For synthetic networks, we report accuracy, measured using three standard criteria: Normalized Mutual Info (NMI), Adjusted Rand Index (ARI), and Adjusted Mutual Info (AMI). For all three accuracy criteria, we used the implementation provided by the Scikit-learn library \cite{scikit-learn}.

\vspace{-.1in}
\section{Performance Study and Results}\label{sec:results}
\vspace{-.1in}

\begin{figure}[htpb!]
\centering
\includegraphics[]{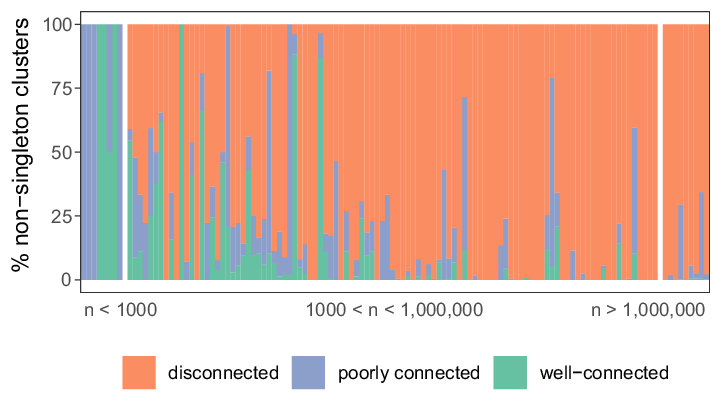}
\caption[Experiment 1: Cluster Connectivity of SBM on Real-World networks]{\textbf{Experiment 1: Cluster Connectivity of SBM on 120 Real-World Networks} Percentage of disconnected, poorly connected, and well-connected clusters are shown for the selected SBM clustering of 120 real world networks. Each colored bar represents a single network, white bars separate the network groups into small, medium, and large. Two of the datasets from the initial set of 122 datasets are not represented here since the selected SBM model returned no non-singleton clusters.}\label{fig:empirical-sbm-connectivity}
\end{figure}
\subsection{Experiments}
We conducted three experiments:

\begin{itemize}
\item Experiment 1: We evaluate the  connectivity profile of SBM clusterings on real-world networks.

\item Experiment 2: We evaluate the impact of returning the connected components of the clusters of SBM clusterings on real-world networks.

\item Experiment 3: We evaluate the impact of our three treatments on clustering accuracy on synthetic networks.

\end{itemize}
 All experiments were performed on the Illinois Campus Cluster \cite{illinois-campus-cluster} with maximum computational resources limit set to 72 hours of runtime, 256 GB of RAM, and 16 cores of parallelism. 

 \vspace{-.1in}
\subsection{Experiment 1: Connectivity of SBMs}

In this experiment, we examined the connectivity profile of clusters generated by SBM. Figure \ref{fig:empirical-sbm-connectivity} shows the edge-connectivity for the selected SBM model (Section \ref{sec:methods-sbm}) on each of the networks, which are sorted from left to right by the number of nodes, which range in size from 11 to 13,989,436 nodes.
Here, red indicates that the cluster is disconnected, blue indicates poorly connected where the edge connectivity is at most $\log_{10}n_c$, where $n_c$ is the number of nodes in the cluster), and green indicates well-connected (i.e., edge cut size more than $\log_{10}n_c$). As we can see, for networks with at most 1000 nodes, clusters are connected, and often well-connected.
Above 1000 nodes, however, the clusters in the selected SBM are very often disconnected, and most clusters are disconnected for most of the networks in the upper half of the size range.

 \vspace{-.1in}
\subsection{Experiment 2: Impact of treatments on real-world networks}

\begin{table}[h]
\centering
\caption[Impact of Treatment on Node Coverage on Real-World Networks]{\textbf{Impact of Treatment on Node Coverage on Real-World Networks} For small, medium, and large network groups, the node coverage (i.e., percentage of  nodes in non-singleton clusters) is shown for the selected SBM before and after treatment.   
On one of the medium networks, WCC ran into a memory error with 256GB of RAM, hence the results for that network are omitted from this calculation. }
\label{tab:chosen-sbm-treatments-node-coverage}
\begin{tabular}{@{}lccc@{}}
\toprule
clustering & Node & Coverage \\
& small  & medium  & large \\
\midrule
Selected SBM & 62\%& 100\%& 100\%\\
Selected SBM - CC& 62\%& 48\%& 45\%\\
Selected SBM - WCC& 55\%& 36\%& 25\%\\
Selected SBM - CM& 55\%& 25\%& 17\%\\
\bottomrule
\end{tabular}
\end{table}

The three post-clustering treatments we apply operate by breaking an input clustering into sub-clusters, and so will change the cluster size distribution, the number of clusters, and even node coverage (i.e., the percentage of nodes in non-singleton clusters). 
Specifically, if in the process singleton clusters are created, then the node coverage, which is calculated based on non-singleton clusters, can also reduce.

We first examine node coverage  (Table \ref{tab:chosen-sbm-treatments-node-coverage}). 
Note that node coverage drops are relatively small on the small networks, but all three treatments produce  large drops in node coverage on the medium and large networks.
The largest drops are for CM, and the smallest drops are for CC (connected components), with WCC in between.
However, even CC produces a large drop in node coverage.
Since node coverage is the percentage of nodes in non-singleton clusters, this drop can only occur because enough nodes are placed in clusters where they do not have any neighbors.  

\begin{figure}[htpb!]
\centering
\includegraphics[]{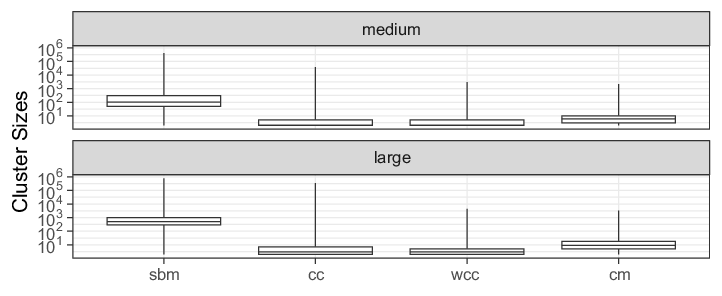}
\caption[Experiment 2: Impact of Treatment on Cluster Sizes of Medium and Large Real-World Networks ]{\textbf{Experiment 2: Impact of Treatment on Cluster Sizes of Medium and Large Real-World Networks}
The distribution of non-singleton cluster sizes  is shown as a boxplot for the selected SBM and its treatments. The y-axis is plotted on a log scale with the whiskers indicating the minimum and maximum cluster sizes in all of the networks in the group.
Both groups and treatments have minimum cluster size of 2 for SBM clusterings whether treated or not, but differ in the medians and maxes, as follows.
Medium group median/max: SBM: 103/403801, SBM+CC: 2/38539, SBM+WCC: 2/2966, SBM+CM: 6/2169.
Large group median/max: SBM:   507/777770, SBM+CC:  3/337018, SBM+WCC:  3/4387, SBM+CM: 9/3258.
}\label{fig:medium-large-cluster-size-boxplot}
\end{figure}

We next examine the impact on cluster size distribution (Figure \ref{fig:medium-large-cluster-size-boxplot}). 
For both medium-sized networks (top) and large networks (bottom),  the median cluster size  before treatment is much larger than the final median cluster size after treatment, and this holds for all three treatments.
Moreover, the majority of clusters are dramatically reduced in size by the treatments.
Even CC, which only modifies the clusters to return connected components,   produces a large impact on the cluster size distribution. Interestingly, the cluster sizes seem to be impacted less when CM treatment is applied compared to CC or WCC, both of which have similar impacts on the cluster sizes regardless of network size.

Finally, we examine the impact on the number of non-singleton clusters (Supplementary Materials).
All three treatments increase the number of clusters, and on average CM produced  the smallest number of non-singleton clusters, CC produced the next smallest, and then  WCC, which produced the most clusters.

 \vspace{-.1in}
\subsection{Experiment 3: Impact of Treatment on Synthetic Networks}
In order to assess the impact of these treatments beyond empirical properties of clusterings, we use synthetic LFR networks with ground truths to capture the effect treatment has on clustering accuracy. We evaluated NMI, ARI, and AMI accuracies on the LFR networks tested.
Recall that some LFR networks had disconnected clusters and so are not part of this evaluation
(i.e., the LFR networks based on CEN clustered using Leiden-CPM  with $r=0.1$ or $r=0.5$, and the wiki\_topcats  clustered using Leiden-CPM with $r=0.5$).   
On the LFR network for cit\_patents with $r=0.5$, WCC treatment could not produce a clustering within our runtime limit of 72 hours when starting with the selected SBM clustering, and so had a ``time-out".

\begin{figure}[htpb!]
\centering
\includegraphics[]{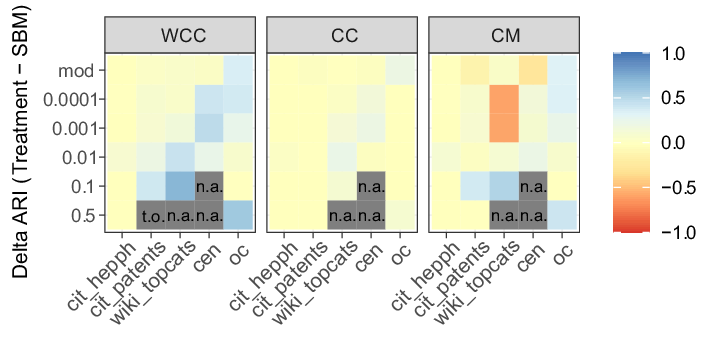}
\caption[Experiment 3: Impact of Treatment on ARI scores of Selected SBM (heatmap)]{\textbf{Experiment 3: Impact of Treatment on ARI Scores of Selected SBM (heatmap).}  
Each LFR network is based on a Leiden clustering  of a real-world network, with the column indicating the real-world network and the row specifying the optimization problem (either modularity or CPM for a given resolution value).  
Blue indicates that post-processing using the corresponding treatment improves ARI accuracy for the clustering method, orange and red  indicate that treatment hurts ARI accuracy, and yellow indicates neutral impact. We use ``n.a.'' to indicate that a network was either not used because of too many disconnected ground-truth clusters or that the LFR software failed to generate the network, and  ``t.o." to indicate that WCC failed to complete within 72 hours.   }\label{fig:lfr-sbm-ari}
\end{figure}

Figure \ref{fig:lfr-sbm-ari} shows through a heatmap that WCC and CC treatments range from neutral (yellow) to beneficial (blue) with respect to the ARI accuracy of SBM clusterings, but WCC improvements are both more frequent and larger than CC improvements. 
In comparison, CM can even be detrimental.
The relative benefit of WCC over CC and CM holds as well for
 NMI and AMI  (Supplementary Materials),
 but for those criteria the impact is generally lessened.
Overall, therefore,  WCC is the preferred treatment for SBM on these networks.

\begin{figure}[htpb!]
\centering
\includegraphics[]{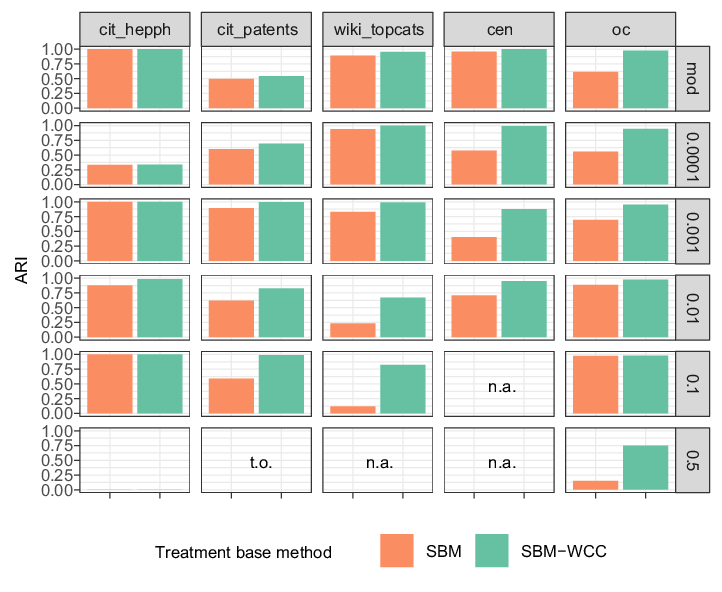}
\caption[Experiment 3: Impact of WCC Treatment on ARI scores of Selected SBM (bar chart)]{\textbf{Experiment 3: Impact of WCC Treatment on ARI Scores of Selected SBM (bar chart).} 
Some LFR networks had too many disconnected ground truth clusters or failed to generate, and so results on these networks are not provided and are marked as ``n.a.". 
WCC on the  LFR network for cit\_patents with Leiden optimizing CPM under $r=0.5$ timed out after 72 hours, and is marked as ``t.o.".  On the LFR network for cit\_hepph with $r=0.5$,  both the selected SBM model and its follow-up WCC yielded 0.0 ARI accuracy.
}\label{fig:lfr-wcc-sbm-ari}
\end{figure}

We explore the impact of WCC in greater detail, noting the ARI accuracy for SBM and the final accuracy for SBM-WCC  (Figure \ref{fig:lfr-wcc-sbm-ari}).
Note that results are not shown (marked as ``n.a.") for some LFR networks (i.e., the 0.5 CEN, 0.1 CEN, and 0.5 wiki\_topcats), 
because they have many disconnected ground truth clusters or failed to generate, as discovered in \cite{park2024}.  On the 0.5 cit\_patents network, WCC treatment could not produce a clustering within our runtime limit of 72 hours when starting with the selected SBM clustering, and so is marked as a time-out (``t.o.").

In every case, SBM-WCC is at least as accurate as SBM.
Moreover, there are many cases where the benefit from WCC treatment is very large (e.g., Open Citations with resolution value 0.5, the Curated Exosome Network with resolution value 0.001, wiki\_topcats with resolution values 0.01 and 0.1). 
Finally, most of the cases where WCC has at most a small positive impact are for the cases where SBM is already highly accurate, with close to 1.00 ARI accuracy, so that there is no room for improvement.

We also examined the models selected by SBM (i.e., degree corrected, non-degree corrected, or planted partition) on the LFR networks. We found that on the LFR networks,  the non degree corrected SBM never yielded the lowest description length, and that typically the planted partition model had the lowest description length  (Supplementary Materials).
We also saw that the lowest description length model resulted in the best NMI/ARI accuracies on the LFR networks tested.

\section{Discussion}
\label{sec:discussion}

\vspace{-.1in}
\subsection{Summary of trends}

As seen in Figures \ref{fig:lfr-sbm-ari} and \ref{fig:lfr-wcc-sbm-ari}, both CC and WCC  post-processing treatments improve accuracy for SBM clusterings  on simulated datasets. CM sometimes improves accuracy but sometimes reduces accuracy, and so is not recommended. Moreover, WCC is particularly beneficial for accuracy.
The impact on cluster size and node coverage produced by even the simple CC technique is significant, and points out interesting questions about SBM clusterings if not followed by CC or WCC treatments.

\vspace{-.1in}
\subsection{Comparison to Prior Work}

In this study, we found that the inner loop within the CM pipeline (which does not remove small clusters) reduced accuracy on occasion, when used with SBM clusterings.
Since the CM pipeline only reduced accuracy for Leiden clusterings when it removed small clusters, this marks a difference between SBM and Leiden clusterings, which merits further investigation.

One key difference between the Leiden algorithm and SBM is that Leiden is guaranteed to produce connected clusters \cite{traag2019louvain} while SBM clearly does not have this guarantee.
Hence the CC treatment of Leiden clusterings would not modify the clustering. 
Nevertheless CC, which only returns the connected components of each cluster, is able to boost the accuracy of SBM clusterings. 

\vspace{-.1in}
\subsection{Impact of the Description Length Formula}

Given that the construction of SBMs within graph-tool produces disconnected clusters, we now consider how the code operates.
Recall that this approach seeks to generate SBMs that optimize the description length, and this is a minimization problem.
Let
\begin{itemize}
    \item $A$ be the adjacency matrix, 
    \item $b$ be the cluster (block) assignment,
    \item $k$ be the degree vector (induced by $A$),
    \item and $e$ be the edge count matrix (induced by $A$ and $b$).
\end{itemize}
In Equation \ref{eq:dcsbm-dl} we provide the formula for the description length of a clustering $b$ of a network given by its adjacency matrix $A$ (i.e., DL(A,b)) under the Degree Corrected (DC) model:
\begin{eqnarray}
    \text{DL}(A, b) = - \log p(A|b, e, k) - \log p(k|b, e) - \log p(b) - \log p(e)
\label{eq:dcsbm-dl}
\end{eqnarray}
Note that the description length is calculated as the sum of various components:  the negative logarithm of the model likelihood (i.e., $- \log p(A|b,e,k)$) and the negative logarithm of each of the priors. 

In our analyses   (Supplementary Materials),  we observed that the model likelihood without priors favors connected clusters returned by the CC treatment. In contrast, certain priors heavily penalize having many clusters, leading to a worse description length for the clustering returned by the CC treatment.

\begin{table}[htpb!]
\centering
\caption[Breakdown of Description Lengths on the \texttt{linux} real-world network]{\textbf{Breakdown of Description Lengths on the \texttt{linux} real-world network} The sum of the values in the first four rows is the value in the last row for all columns. The ratio is SBM(DC)-CC / SBM(DC), so that values less than $1.0$ favor SBM treated by CC  and values greater than $1.0$ favor untreated SBM. Bold text indicates the preferred clustering for each row where lower values are more desirable. Here, $A$ is the adjacency matrix, $b$ is the clustering, $k$ is the degree vector, and $e$ is the edge count matrix.}
\label{tab:dl-example}
\begin{tabular}{lrrr}
\toprule
Quantity & SBM(DC) & SBM(DC)-CC & Ratio \\
\midrule
$-\log p(A|b, e, k)$ & 699228.26 & \textbf{315644.88} & 0.45 \\
$-\log p(k|b, e)$ & 95737.43 & \textbf{45066.47} & 0.47  \\
$-\log p(b)$ & \textbf{147018.92} & 256817.11 & 1.75 \\
$-\log p(e)$ & \textbf{50786.40} & 1584554.98 & 31.20 \\
\midrule
DL$(A, b)$ & \textbf{992771.01} & 2202083.44 & 2.22 \\
\bottomrule
\end{tabular}
\end{table}
We provide an example of this phenomenon on a real-world network, \texttt{linux}, in Table \ref{tab:dl-example}; in the Supplementary Materials, we show that 
the trends observed on this network are also observed in the other real-world networks.

One clustering is from SBM(DC), the degree corrected SBM output, while the second clustering is from SBM(DC)-CC, the result of running the CC treatment on the degree corrected SBM output. 
Note that the SBM(DC) clustering has a lower description length than  SBM(DC)-CC, and hence the untreated SBM clustering is the preferable clustering with respect to the minimization of the description length under the degree corrected model. 
However, although $-\log p(A|b, e, k)$ and $-\log p(k|b, e)$ for the SBM(DC)-CC clustering are lower, $-\log p(b)$ and $-\log p(e)$ for SBM(DC)-CC clustering are higher, and by a larger magnitude,  and hence offset the first two quantities.
Moreover, between these two priors, the $- \log p(e)$ component has the bigger impact on this outcome.
Furthermore, if this component had been {\em not included} in the summation, then SBM(DC)-CC would have a lower description length, and would have been favored.

We examined the other 102 networks that had selected DC as the model.  For all of these, the $-\log p(e)$ component strongly favored the untreated SBM over the treated SBM. We then examined whether  removing the $-\log p(e)$ component of the summation of the description length for both treated and untreated SBM models, to see which model would have been returned.  We found that for 80 of the 103 networks in total, removing this component of the summation would have resulted in the treated SBM model having a lower description length than the untreated model.
Thus, this specific component of the summation accounts for 77.7\% of the cases where the untreated SBM is favored over the treated SBM.

Examining the formula for  $-\log p(e)$ (Supplementary Materials), we see that it increases quickly as the number of clusters increases.
This explains why clusterings with a larger number of clusters (such as are produced by running CC) have larger description lengths, and hence are less favored.

\section{Conclusion}\label{sec:conclusion}
We observe that clustering using SBMs is prone to producing disconnected clusters, with the frequency of disconnected clusters increasing as the network size grows. 
We show that a simple technique, CC, which returns the connected components of the clusters, can be used to modify an SBM clustering and improve clustering accuracy on synthetic networks.  The application of either the Connectivity Modifier or the simpler WCC technique modify the clustering to ensure well-connectedness. Both CM and WCC  tend to improve accuracy on synthetic networks--but WCC has the strongest improvements in our simulation study. 
In summary, we present simple techniques for improving SBM clusterings with respect to connectivity.

\bibliographystyle{spmpsci} 
\bibliography{main} 
\end{document}